\documentclass[superscriptaddress,twocolumn,showpacs,preprintnumbers,amsmath,amssymb]{revtex4}

\usepackage{graphicx}
\usepackage{dcolumn}
\usepackage{bm}
\usepackage{color}
\usepackage{ulem}
\usepackage{epsfig}

\begin{document}
\preprint{APS/123-QED}

\title{Minimum reflection channel in amplifying random media }

\author{Seng Fatt Liew}
\author{Hui Cao}
\email{hui.cao@yale.edu}
\affiliation{Applied Physics Department, Yale University, New Haven CT 06520, USA.}

\date{\today}

\begin{abstract}
We present a numerical study on the minimum reflection channel in a disordered waveguide and its modification by coherent amplification of light. 
The minimum reflection channel is formed by destructive interference of quasi-normal modes at the front surface of the random medium. 
While the lowest reflection eigenvalue increases with gain in most random realizations, the minimum reflection channel can adjust its modal composition to enhance the destructive interference and slow down the growth of reflectance with gain. 
Some of the random realizations display a further reduction of the minimum reflectance by adding optical gain. 
The differential amplification of the modes can make their destructive interference so effective that it dominates over the amplitude growth of the modes, causing the reflectance to drop with gain. 
Therefore, the interplay between interference and amplification makes it possible to further minimize light reflection from a strong scattering medium by introducing optical gain. 

\end{abstract}

\maketitle

\section{\label{sec:level1}Introduction}

\indent Due to multiple scattering of light, disordered media such as paper, paint or biological tissue have very high reflection and look opaque. 
However, wave interference effect may diminish the reflectance by creating highly transmitting channels called “open channels” \cite{dorokhov1,dorokhov2,Mello88, Nazarov94}. 
The recent developments of adaptive wavefront shaping and phase recording techniques in optics have enabled the coupling of incident light to these open channels \cite{vellekoop_OL, changhueiyang_NP, vellekoop_PRL, Pendry_view, choi_PRB, choi_OE, shi_PRL12, choi_NP, Davy_OE13, sebastien13, sebastien_PRL, mosk_NP,kim_OL13,liew_PRB14,Genack_NC15}. 
The open channels can greatly enhance light transmission through scattering media, that will have a profound impact in a wide range of applications from deep tissue imaging and laser surgery to spectroscopy and opto-genetics \cite{sebastien_NC,Lagendijk_PRL11, Katz_NP, Gigan_NP, Mosk_NP12,Park2012spectral,Small2012control,Piestun,Mertz, Yu_review, Kubby, Park_OptoGen}. 
So far most experimental studies of high transmission channels rely on the optical access to both sides of the scattering media, which is not practical in realistic situations. 
It is known that there is an one-to-one correspondence between the transmission eigenchannels and reflection eigenchannels in turbid media without absorption. 
Thus the information about the transmission channels may be obtained from the reflection measurements, which are conducted on the input side of the samples and thus less invasive \cite{Park_OC15}. 
For example, by reducing reflection, one can couple light into the minimum reflection channel, which corresponds to the maximum transmission channel. \\

\indent The correspondence between transmission and reflection holds only when there is no absorption. 
However, absorption exists in many material systems and is known to have a significant impact on the mesoscopic transport of light \cite{John_PRL84,Yurkevich_PRL94,zhang_PRB95,beenakker_PhysA,Heinrichs_PRB97,brouwer,Datta_PRB99,chabanov_Nat00,Lisyansky_PRB01,yamilov_OE13,Zhang_PRB13,liew_PRB14,liewOE15}. 
Our recent studies show that when strong absorption is introduced uniformly across a diffusive system, the maximum transmission channel turns into quasi-ballistic \cite{liew_PRB14}. 
In case the absorption is distributed non-uniformly in space, a high transmission channel may redirect the energy flow to circumvent the strong absorbing regions \cite{liewOE15}.
Because light absorption can also reduce reflection, the minimum reflection channel no longer corresponds to maximum transmission channel \cite{liew_PRB14}. 
Usually the optical gain has the opposite effects to the absorption, and there have been extensive studies on the effects of coherent amplification on light propagation in random media \cite{Zyuzin_94, Kumar_PRB94,zhang_PRB95, Zyuzin_PRE95, Wiersma_PRL95, beenakker_PRL96,Burkov_JETP96, Burkov_PRB97, beenakker_PhysA, Heinrichs_PRB97,Freilikher_PRB97,Jayannavar_PRB97, soukoulis_PRB99,Datta_PRB99,Kumar_PRB00,Kumar_PRB00v2,Kumar_PhysE01,Yamilov_PRE04,Yamilov_PRB05,Yamilov_PRE06,Yamilov_PRB10}. 
However, it is not yet clear how optical amplification would modify the transmission and reflection channels. 
Intuitively, one would expect both transmittance and reflectance to increase with gain due to light amplification. 
Such an expectation does not take into account the interference of multiply scattered light, which plays a dominant role in enhancing transmission and suppressing reflection. 
As illustrated in the previous studies on the random lasers, coherent amplification can enhance the interference effects in random media, which may lead to unexpected behavior \cite{CaoWRM, CaoJPA}. \\

\indent In this paper, we present a numerical study to answer the following questions: what is the impact of light amplification on the minimum reflection channel? 
Is it possible to further reduce reflectance of a strong scattering medium by adding optical gain to it? 
How will the interference effect that underlies the formation of minimum reflection channel be modified by coherent amplification?  
We numerically calculate the minimum reflection channels in disordered waveguides with optical gain. 
We find that the minimum reflection channel is formed by destructive interference of quasi-normal modes at the front surface of the random system. 
In most of the disordered waveguide configurations, adding gain to the system causes the reflectance to increase, but the increment can be slowed down by adjusting the input wavefront to enhance the destructive interference effect. 
In some of the disordered waveguides, the enhancement of the destructive interference is so strong that the reflectance drops with increasing gain. 
This result illustrates that in the random media with gain, the coherent phenomena (due to interference of excited modes) may dominate over the incoherent phenomena (due to growth of mode amplitudes). 
Thus optical gain provides an additional degree of freedom for coherent control of mesoscopic transport of light.  \\

\indent This paper is organized as follows. 
Section II presents our numerical model of two-dimensional (2D) disordered waveguides with gain. 
In section III we demonstrate the modification of the minimum reflection eigenvalue by gain. 
In section IV we conduct a modal analysis of the reflected light.
In Section V, the interference effects among the quasi-normal modes that constitute the minimum reflection channels are investigated. 
Section VI illustrates how the modal interference can further reduce the minimum reflection in the presence of gain. 
Section VII is the conclusion.


\section{\label{sec:level2} Disordered waveguide with gain}

\begin{figure}[htbp]
\centering
\includegraphics[width=\linewidth]{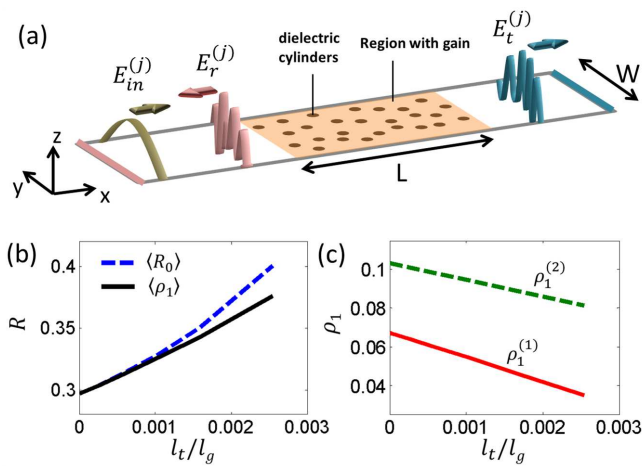}
\caption{(Color online) (a) Schematic of the 2D disordered waveguide we simulate. Dielectric cylinders are placed randomly in a waveguide with perfect-reflecting sidewalls. An electric field $E_{in}^{(j)}$ is launched from the left end of the waveguide, and scattered by the cylinders. The reflected electric field $E_r^{(j)}$ is probed at the left end, and the transmitted electric field $E_t^{(j)}$ at the right end. Perfectly-matched-layers are placed at both open ends to absorb the transmitted and reflected waves. 
	(b) The ensemble-averaged reflectance $R_0$ for input wavefront corresponding to the minimum reflection channel without gain (dashed line) and the minimum reflection eigenvalue $\rho_1$ at each gain level (solid line) as a function of optical gain $l_t/l_g$. 
	(c) The minimum reflection eigenvalue $\rho_1$ drops with increasing gain $l_t/l_g$ in two of the disordered waveguides in the ensemble.}
\label{Fig1}
\end{figure}

Our numerical model consists of a 2D disordered waveguide, shown schematically in Fig. \ref{Fig1}(a). 
Dielectric cylinders with refractive index $n_s=2.5$ and radius $r_c=0.1\lambda$ are randomly positioned inside the waveguide with perfectly reflecting sidewalls. 
The average distance between adjacent cylinders is $a=0.87 \lambda$, giving an area filling fraction of 0.04 for the dielectric in air. 
The wavelength of the input light is chosen to be away from the Mie resonances of individual cylinders, so that the scattering properties of the random system do not vary strongly with frequency. 
The incident light enters the waveguide from an open end and is scattered by the cylinders in the $x$-$y$ plane. 
The light transmitted through or reflected from the random array is absorbed by the perfectly-matched-layers located at both ends of the waveguide. \\

\indent The probe light is transverse-magnetic (TM) polarized, with the electric field parallel to the cylinder axis ($z$-axis). 
The width of the waveguide is $W = 10.3\lambda$, and the number of guided modes in the empty waveguide is $N = 2W/\lambda  = 20$. 
The length of the random array of cylinders is $L = 20.2\lambda$. 
The transport mean free path is $l_t = 0.07L$ and the localization length is $\xi = (\pi /2) N l_t = 2.3L$. 
The system is in the diffusion regime but not far from the localization threshold. 
Thus the transport displays a large fluctuation from one random configuration to another. 
Within the same statistical ensemble there are random realizations that are closer to or further away from the localization transition. 
This allows us to study the diverse behavior in the same ensemble.\\

\indent Usually optical gain exists either inside the scattering particles or in the background material that hosts the particles. 
The contrast in the imaginary part of the refractive index causes additional scattering \cite{wu_JOSAB07}, which is avoided here by introducing gain to both the scatterers and the host material. 
More specifically, the optical gain is introduced uniformly across the scattering region [highlighted in Fig. \ref{Fig1}(a)] by adding a constant imaginary part -$\gamma$ to the refractive index $n = n_0 - i \gamma$, where $n_0$ is the real part of refractive index without gain.  
The gain length is $l_g = 1/(2k\gamma)$, where the wavevector $k = 2 \pi/ \lambda$. 
When the gain length $l_g$ reaches the average path length of light in a 2D diffusive waveguide of length $L$, $2 L^2 / l_t$, the diffusive amplification length $l_{amp} = \sqrt{l_tl_g/2}$ becomes equal to $L$. 
$L = l_{amp}$ corresponds to the lasing threshold of a diffusive random laser \cite{CaoWRM}, above which nonlinear gain saturation must be taken into account. 
To stay in the linear gain regime, we make sure the amount of gain is below the lasing threshold $L < l_{amp}$.   
We check individual configuration and verify that all of the random systems stay below lasing threshold even in the presence of fluctuation. 
The spontaneous emission and its amplification are ignored in the calculation below. \\

\indent To construct the reflection matrix $r$ of the disordered waveguide, we use the guided modes in the empty waveguide (without scatterers) as the basis. 
The electromagnetic field inside the disordered waveguide was calculated using the finite-difference frequency-domain method \cite{comsol_42a}.
We launch a guided mode $E_{in}^{(j)}(y)$ from the input end $(x = 0)$, calculate the reflected wave and decompose it by the empty waveguide modes at $x = 0$, $E_r^{(j)}(y) = \sum\limits_{i=1}^{N}r_{ij} E_{in}^{(j)}(y)$. 
The coefficient $r_{ij}$ relates the field incident into the waveguide mode $j$ to the field reflected to the waveguide mode $i$. 
After repeating this procedure for $j = 1, 2, ... N$, we obtain all the elements $r_{ij}$ for the reflection matrix $r$. 
Similarly, the transmission matrix $t$ is constructed by computing the transmitted waves $E_t^{(j)}(y)$ at the output end $x = L$. \\

\section{\label{sec:level3} Modification of minimum reflection by gain}

\indent A singular value decomposition of the reflection matrix $r$ gives
\begin{eqnarray}
r = U\  \Sigma \ V^{\dagger} \ , 
\end{eqnarray} 
where $\Sigma$ is a diagonal matrix with non-negative real numbers $\sqrt{\rho_n}$, $\rho_n$ is the eigenvalue of $r^{\dagger} r$, $\rho_1 < \rho_2 < \rho_3 \ ... < \rho_N$. 
$U$ and $V$ are $N \times N$ unitary matrix, $V$ maps input waveguide modes to eigenchannels of the disordered waveguide, and $U$ maps eigenchannels to output waveguide modes. 
The column vectors in $V$ ($U$) are orthonormal and are called input (output) singular vectors. 
The input singular vector corresponding to the lowest reflection eigenvalue $\rho_1$ couples to the minimum reflection eigenchannel, its elements represent the complex coefficients of the waveguide modes that combine to achieve minimum reflection from the random medium. 
Similarly, the transmission eigenvalues $\tau_n$ where $\tau_1 > \tau_2 > \tau_3 ... > \tau_N$ is obtained from singular value decomposition of the field transmission matrix $t$. \\

\indent In a passive system without gain or loss, there is an one-to-one correspondence between the transmission and reflection eigenchannels, $\tau_n + \rho_n$ = 1. 
The minimum reflection channel has the same input wavefront as the maximum transmission channel. 
Since our system is close to the localization threshold, some of the random configurations have the maximum transmission eigenvalue $\tau_{1}$ smaller than one. 
Therefore, the ensemble-averaged maximum transmission eigenvalue $\langle \tau_{1} \rangle = 0.7$ and the minimum reflection eigenvalue $\langle \rho_1 \rangle = 0.3$. \\

\indent We fix the input wavefront to be the minimum reflection channel and then introduce optical gain into the system. 
The ensemble-averaged reflectance $\langle R_{0} \rangle$ increases due to light amplification [dashed line in Fig. \ref{Fig1}(b)]. 
We also compute the reflection matrix at each gain level, and from the singular value decomposition we obtain the ensemble-averaged minimum reflection eigenvalue $\langle \rho_{1} \rangle$, which is plotted as a function of $l_t/l_g$ [solid line in Fig. \ref{Fig1}(b)]. 
The increment of $\langle \rho_{1} \rangle$ is slower than $\langle R_{0} \rangle$,  indicating that the minimum reflection channel with gain deviates from that without gain. \\

\indent Surprisingly, we observe that in about 10$\%$ of the disordered waveguides in the ensemble, the minimum reflection eigenvalue $\rho_{1}$ is further reduced by adding gain to the system. 
Figure \ref{Fig1}(c) shows two of these disordered waveguides, where $\rho_{1}$ drops with increasing gain $l_t/l_g$. \\

\section{\label{sec:level4} Modal analysis of reflected light}

To interpret the formation of minimum reflection channel, we resort to the quasi-normal modes, which represent the resonances of an open system. 
The interference between quasi-normal modes has been investigated previously to explain coherent transport of light in random media \cite{genack_nature, genack_JMP, liew_PRB14, modalMakeup}. 
Below we will consider the contribution of quasi-normal modes to the light reflected from the disordered waveguide. 
There are two types of quasi-normal modes: (i) the outgoing modes $u_m$ - the eigenfunctions of the Maxwell's equations which satisfy the boundary conditions that there are only outgoing waves to the infinity; (ii) the incoming modes $v_m$  - the eigenfunctions for the boundary conditions of only incoming waves from infinity.  
In a passive system (without gain or loss), the two types of eigenfunctions are related by $u_m = v_m^*$. 
We use the commercial program COMSOL \cite{comsol_42a} to compute the quasi-normal modes in the disordered waveguide. 
The spatial field distributions of the modes remain the same when gain is introduced uniformly across the system. \\

\indent With light incident onto the random waveguide, both types of the quasi-normal modes are excited, and the electric field distribution inside the disordered medium is  decomposed by (see appendix). 
\begin{equation}
E(x,y) = \sum_m a_m u_m(x,y) + \sum_m b_m v_m(x,y).
\end{equation} 
The reflected field at the front surface ($x=0$) of the disordered waveguide can be expressed by the outgoing modes as
\begin{equation}
E_r(y) = \sum_m a_m u_m(0,y).
\end{equation}  
The contribution of the $m$-th mode to the reflectance $R = \int |E_r(y)|^2 dy $ depends on its overlap with the reflected field $E_r$, namely, 
\begin{equation}
\sqrt{R} = \sum_m a_m \int E_{w}^*(y) \, u_m(0,y) dy = \sum_m \alpha_m, 
\end{equation}
where $E_{w}(y) \equiv E_r(y)/\sqrt{R}$ is the normalized reflected field, and $\alpha_m$ represents the contribution from the $m$-th mode.
$\alpha_m$ is a complex number $\alpha_m = |\alpha_m| e^{i \theta_m}$, where the phase $\theta_m \in [-180^{\circ}, 180^{\circ}]$ determines its interference with other modes. 
The reflectance can be written as
\begin{eqnarray}
R &=&\left\vert\sum_m \alpha_m\right\vert^2 \nonumber\\
 &=& \sum_m \lvert\alpha_m\rvert^2 + \sum_{i,j, j\neq i} \lvert\alpha_i\rvert \, \lvert\alpha_j\rvert \, \cos(\theta_i - \theta_j) \nonumber \\
 &=& R_i + R_c. \nonumber
\end{eqnarray} 
where $R_i \equiv \sum_m \lvert\alpha_m\rvert^2$ is the incoherent sum of the modal contributions to the reflectance $R$, and $R_c \equiv \sum_{i,j, j\neq i} \lvert\alpha_i\rvert \, \lvert\alpha_j\rvert \, \cos(\theta_i - \theta_j)$ is the interference term that depends on the relative phases of the modes, i.e, $\cos(\theta_i-\theta_j) > 0$ gives constructive interference, and $\cos(\theta_i-\theta_j) < 0$ the destructive interference. 
The ratio $C \equiv R / R_i = 1 + R_c/R_i$ quantifies the effect of modal interference on the reflection. 
For a random input wavefront, the phase difference between $\alpha_m$ are randomly distributed in  $[-180^{\circ}, 180^{\circ}]$ , thus $R_{c} \approx 0$ and $C \approx 1$. 

\section{\label{sec:level5} Destructive interference of modes to minimize reflection}
\begin{figure}[htbp]
	\centering
	\includegraphics[width=\linewidth]{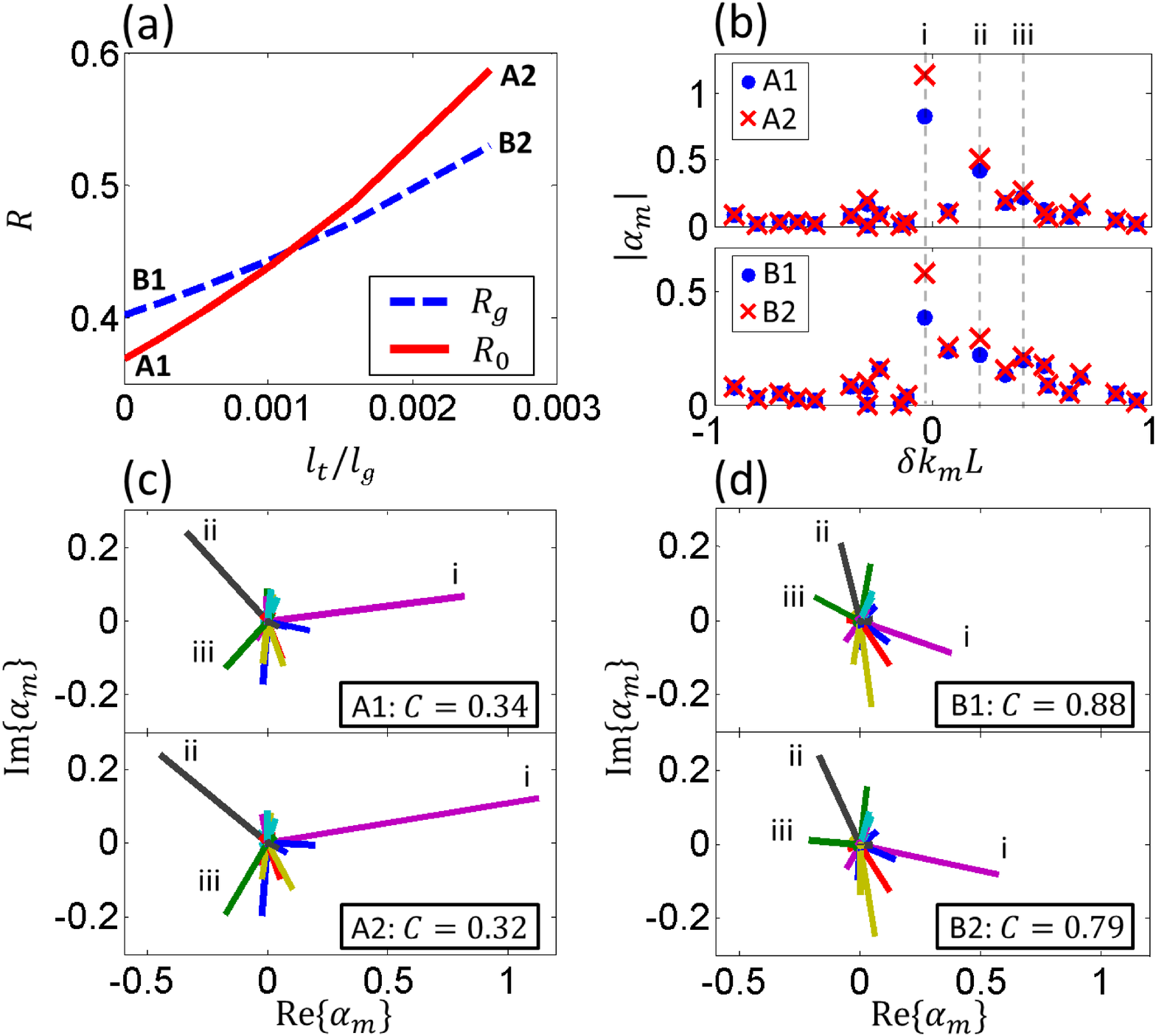}
	\caption{(Color online) Formation of the minimum reflection channel by destructive interference of quasi-normal modes. (a) Solid (dashed) line shows the evolution of reflectance $R_{0}$ ($R_{g}$) with gain $l_t / l_g$ when the input wavefront is set to that of the minimum reflection channel at $l_t/l_g = 0$ ($l_t/l_g = 0.0025$). The crossing of the curves reveals that the optimal wavefront for the minimum reflection channel changes with gain. (b) Amplitudes of the contributions of quasi-normal modes to the reflectance $|\alpha_m|$ at $l_t/l_g = 0$ (A1, B1) and $l_t/l_g = 0.0025$ (A2, B2) with the input wavefront equal to that of the minimum reflection channel at $l_t/l_g = 0$ (A1, A2) and that of $l_t/l_g = 0.0025$ (B1, B2), respectively. The horizontal axis $\delta k_m = k_m-k$ is the difference between the $m$-th mode frequency $k_m$ to the input light frequency $k$. Three modes in the vicinity of $k$, labeled i-iii, have major contributions to the reflected light, and their amplitudes grow with gain at a different rate. (c-d) Individual quasi-normal modes' contributions $\alpha_m$ are plotted in the complex plane to show their relative phases in four cases of A1, A2, B1 and B2. The degree of destructive interference of the modes is characterized by the value of $C$ given in each panel. The three major modes, i-iii, in A1 and A2 interfere destructively to minimize the reflection. The destructive interference effect, which is weaker in B1, is enhanced by gain in B2. \\ }
	\label{Fig2}
\end{figure}

\indent Next we apply the modal analysis to the minimum reflection channel. 
As an example, let us consider a typical disordered waveguide, whose minimum reflection eigenvalue $\rho_1$ is close to the ensemble average and it increases with gain. 
We fix the input wavefront to that of the minimum reflection channel in the absence of gain, and calculate the reflectance $R_{0}$ when gradually increasing gain. 
As shown by the solid line in Fig. \ref{Fig2}(a), $R_{0}$ increases from 0.37 at $l_t/l_g = 0$ (labeled A1) to 0.59 at $l_t/l_g = 0.0025$ (labeled A2). 
The top panel in Fig. \ref{Fig2}(b) plots the amplitude $|\alpha_m|$ of different modes' contribution to the reflection as a function of the difference between the mode center frequency $k_m$ and the incident light frequency $k$, $\delta k_m L = (k_m-k) L$.  
The minimum reflection channel consists mainly of three modes, labeled i-iii, in the vicinity of $k$, with mode i being the most dominant one. 
\\
\\
\\
\begin{table}[pb!]
	\setlength{\tabcolsep}{15pt}
	\caption{Values of reflectance $R$, the incoherent sum of modal contributions $R_i$, the interference term $R_c$ in the cases of A1, A2, B1 and B2 in Fig. \ref{Fig2}(a), as well as their ratio.} 
	\centering 
	\begin{tabular}{c   c   c   c} 
		\hline
		\hline
		\empty & $R$ & $R_i$ & $R_c$ \\ 
		\hline 
		A1 & 0.37 & 1.07 & -0.71  \\ 
		\hline
		A2 & 0.59 & 1.83 & -1.24 \\
		\hline
		A2/A1 & 1.59 & 1.71 & 1.75 \\ 
		\hline 
		B1 & 0.40 & 0.46 & -0.05  \\ 
		\hline
		B2 & 0.53 & 0.71 & -0.19 \\
		\hline
		B2/B1 & 1.33 & 1.54 & 3.80 \\
		\hline 
		\hline
	\end{tabular}
	\label{Stable1} 
\end{table}
\indent To illustrate the modal interference at the front surface of the disordered waveguide, we plot $\alpha_m$ in the complex plane in Fig. \ref{Fig2}(c) for the reflected light with and without gain. 
At $l_t/l_g = 0$, the top panel of Fig. \ref{Fig2}(c) shows that mode i interferes destructively with mode ii and mode iii, as their phase differences fall in 90$^\circ$-180$^\circ$, leading to $R_c =-0.71 < 0$ and $C=$ 0.34 $<$ 1 (Table I). 
Thus the destructive interference minimizes the reflected light intensity. 
When gain is added to the system, the amplitudes $|\alpha_m|$ for modes i-iii increase due to light amplification [A2 in the top panel of Fig. \ref{Fig2}(b)], but they continue to interfere destructively with $C$ = 0.32 at $l_t/l_g = 0.0025$ [bottom panel of Fig. \ref{Fig2}(c)]. 
Both the incoherent sum $R_i$ and the interference term $R_c$ have a similar amplitude growth with gain, leading to an increment of the reflectance $R_0$ [Table \ref{Stable1}]. \\

\indent In the above analysis, the input wavefront is set to that of the minimum reflection channel without gain, which may not be optimal when there is gain.  
We calculate the minimum reflection channel from the reflection matrix in the presence of gain, and find the minimum reflection eigenvalue $\rho_1$ increases slower with gain than $R_0$. 
For example, at $l_t/l_g = 0.0025$, $\rho_1 = 0.53$ [B2 in Fig. \ref{Fig2}(a)], while $R_0 = 0.59$ (A2), indicating the input wavefront can be further optimized in the presence of gain to minimize reflection. 
For comparison, we fix the input wavefront to that of the minimum reflection channel at $l_t/l_g = 0.0025$, and calculate the reflectance $R_g$ when gradually reducing the gain to 0. 
As plotted by the dashed curve in Fig. \ref{Fig2}(a), $R_{g}$ is higher than $R_0$ at $l_t/l_g = 0$ (labeled B1),  but it grows at a lower rate with gain than $R_0$.  
The modal analysis [bottom panel of Fig. \ref{Fig2}(b)] reveals that the minimum reflection channel evolves with gain to slow down the growth of reflection due to amplification. 
Compared to $R_0$ [top panel of Fig. \ref{Fig2}(b)], more modes are excited and contribute to the reflectance $R_g$. 
Without gain, the destructive interference of modes in $R_g$ is less efficient, the interference term $R_c = -0.05$ is very small, and  the ratio $C$ = 0.88 is closer to 1 [top panel of Fig. \ref{Fig2}(d)]. 
With the introduction of optical gain, the modal interference becomes more destructive, $R_c = -0.19$ and $C = 0.79$ at $l_t/l_g = 0.0025$ [bottom panel of Fig. \ref{Fig2}(d)]. 
As listed in Table I, the magnitude of the interference term $|R_c| = -R_c$ increases by a factor of 3.8, while the incoherent sum, $|R_{i}| = R_i$, only by a factor of 1.54. 
The enhancement of the destructive interference effect makes the growth of $R_g$ slower than $R_0$. \\

\section{\label{sec:level6} Reduction of minimum reflection by gain}
\indent The above section shows that the minimum reflection channel adjusts its modal composition to enhance the destructive interference effect in the presence of gain. 
This slows down the increase of the minimum reflection eigenvalue with gain, which is a typical behavior for most of the disordered configurations. 
However, in some of the disordered waveguides, e.g, the two shown in Fig. \ref{Fig1}(c), the minimum reflection eigenvalue is further reduced with the addition of gain. 
Below we explain this behavior by the interference of the constituent modes at the front surface of the disordered waveguide. 
\begin{figure}[ht!]
	\centering
	\includegraphics[width=\linewidth]{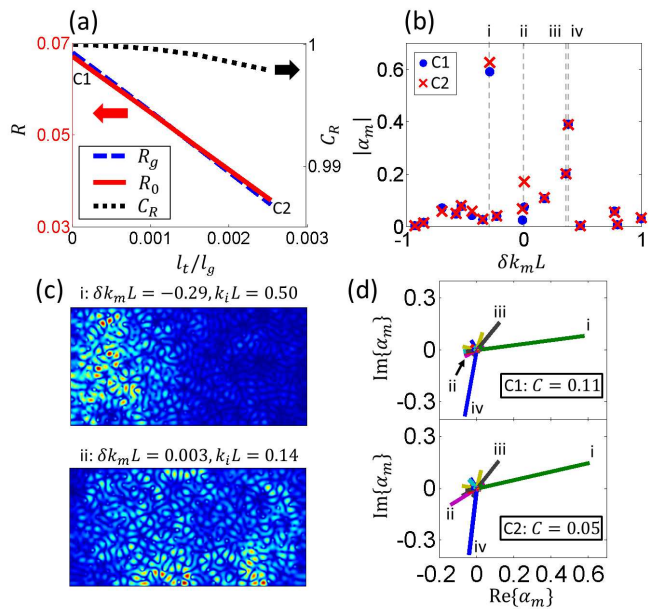}
	\caption{(Color online) (Color online) Further reduction of the minimum reflection by gain. (a) The correlation $C_R$ between the input wavefront for the minimum reflection channel with gain to that without gain, showing they are highly correlated. The reflectance $R_0$ with the input wavefront equal to that of the minimum reflection channel at $l_t / l_g = 0$ almost coincides with $R_g$ with input wavefront set to that of the minimum reflection channel at $l_t / l_g = 0.0025$. Both drops with increasing gain $l_t / l_g$. (b) Amplitudes of modes' contributions $|\alpha_m|$ to the reflection without gain [C1 in (a)] and with gain [C2 in (a)], showing the major contributions from four modes labeled i-iv. (c) Spatial distribution of electric field amplitude $|E_z(x,y)|$ for mode i and ii in (b). (d) Without gain $l_t / l_g = 0$, the modes interfere destructively to form the minimum reflection channel, $C$ = 0.11 [top panel]. With gain $l_t / l_g = 0.0025$, the destructive interference of modes is further enhanced, $C$ is reduced to 0.05 [bottom panel]. }
	\label{Fig3}
\end{figure}
\\
\begin{table}[h!]
	\setlength{\tabcolsep}{15pt}
	\caption{Values of reflectance $R$, the incoherent sum of modal contributions $R_i$, the interference term $R_c$ in the cases of C1 and C2 in Fig. \ref{Fig3}(a), as well as their difference. } 
	\centering 
	\begin{tabular}{c   c   c   c} 
		\hline
		\hline
		\empty & $R$ & $R_i$ & $R_c$ \\ 
		\hline 
		C1 & 0.067 & 0.60 & -0.54  \\ 
		\hline
		C2 & 0.035 & 0.68 & -0.65 \\
		\hline
		C2$-$C1 & -0.032 & 0.08 & -0.11 \\ 
		\hline 
		\hline
	\end{tabular}
	\label{Stable2} 
\end{table}
\indent First we study the random waveguide with the minimum reflection eigenvalue labeled $\rho_1^{(1)}$ in Fig. \ref{Fig1}(c). 
Figure \ref{Fig3}(a) shows that the input wavefront for the minimum reflection channel with gain is highly correlated to that without gain. 
This means the minimum reflection channel is barely modified in the presence of gain. 
Nevertheless, the reflectance drops with increasing gain, even when the input wavefront is fixed to that of the minimum reflection channel in the absence of gain.  
The modal decomposition reveals that the minimum reflection channel without gain is dominated by one mode, labeled mode i in Fig. \ref{Fig3}(b), which is slightly detuned from the frequency of input light. 
Another mode, labeled mode ii, has the frequency closest to the input light but its contribution to the reflected light is much less. 
Such difference can be understood from the mode's spatial field profile $|E_z(x,y)|$ and spectral width $k_i$ (the imaginary part of the eigenfrequency). 
As shown in the top panel of Fig. \ref{Fig3}(c), mode i is concentrated spatially close to the front surface of the disordered waveguide, resulting in a large leakage rate. 
Consequently, mode i has a broad spectral width that exceeds the detuning of its center frequency from the input frequency, $k_i > \delta k_m$. 
The spectral overlap of mode i with the input light leads to an efficient excitation of mode i by the input light, and mode i's proximity to the front surface of the waveguide enhances its contribution to the reflected light. 
In contrary, mode ii penetrates deeper into the waveguide and has a smaller leakage rate, thus its contribution to the reflected light is much less [bottom panel of Fig. \ref{Fig3}(c)]. \\

\indent With the introduction of gain, mode ii experiences stronger amplification than mode i due to its lower leakage rate, and its contribution to the reflection grows faster than mode i [Figs. \ref{Fig3}(b)]. 
As shown in top panel of Fig. \ref{Fig3}(d), mode i and mode ii have a phase difference of $\sim$ 180$^\circ$, so they interfere destructively.
When there is no gain, the relative large difference in their amplitudes makes their interference ineffective. 
In the presence of gain, the imbalance of their amplitudes is reduced, thanks to the faster growth of mode ii, and their destructive interference becomes more effective. 
This means the differential amplification of individual modes can enhance the degree of destructive interference, as confirmed by the reduction of $C$ from 0.11 at $l_t/l_g = 0$ to 0.05 at $l_t/l_g = 0.0025$ [Fig. \ref{Fig3}(d)]. 
More quantitatively, the interference term $R_c$ decreases from -0.54  at $l_t/l_g = 0$ to -0.65 at $l_t/l_g = 0.0025$, while the incoherent sum $R_i$ increases from 0.60 to 0.68  (Table II). 
The increment of $R_i$ is not sufficient to compensate for the reduction of $R_c$, thus the reflectance $R$ decreases with gain. \\

\begin{figure}[ht!]
	\centering
	\includegraphics[width=\linewidth]{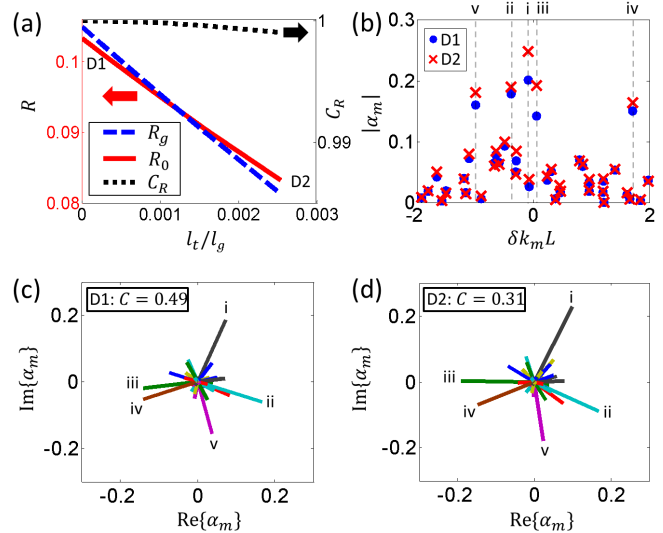}
	\caption{(Color online) Another example showing the minimum reflection eigenvalue decreases with gain. (a) The correlation $C_R$ between the input wavefront for the minimum reflection channel with gain to that without gain, showing they are highly correlated. The reflectance $R_0$ for the input wavefront fixed to that of the minimum reflection channel without gain and the reflectance $R_g$ for input wavefront fixed to that of the minimum reflection channel with gain ($l_t / l_g = 0.0025$) both decrease with increasing gain. (b) Amplitudes of modes' contributions $|\alpha_m|$ to the reflection without gain [D1 in (a)] and with gain [D2 in (a)], showing the major contributions from five modes labeled i-v. (c) Without gain ($l_t / l_g = 0$), the modes interfere destructively to form the minimum reflection channel, $C$ = 0.49. (d) With gain $l_t / l_g = 0.0025$, the destructive interference effect is further enhanced and $C$ is reduced to 0.31.  }
	\label{Fig4}
\end{figure}

\indent Next we investigate the second configuration of the disordered waveguide shown in Fig. \ref{Fig1}(c). 
Similar to the first one, the input wavefront for the minimum reflection channel with gain is nearly identical to that without gain [Fig. \ref{Fig4}(a)]. 
The reflectance with the input wavefront fixed to that of the minimum reflection channel without gain $R_0$ is almost the same as the reflectance with the input wavefront set to that of the minimum reflection channel with gain $R_g$, they both decrease as gain increases. 
The modal decomposition reveals that there are more modes contributing to the reflected light [Fig. \ref{Fig4}(b)], compared to the previous two cases [Figs. \ref{Fig2} and \ref{Fig3}]. 
These modes interfere destructively to form the minimum reflection channel when there is no gain [Fig. \ref{Fig4}(c)], and the destructive interference effect is further enhanced by adding gain to the system [Fig. \ref{Fig4}(d)]. 
Hence, the reduction of the reflectance due to enhanced destructive interference effect exceeds the increment due to the growth of the modes' amplitudes with gain, leading to a further reduction of the minimum reflectance. \\
\\
\\
\\
\begin{table}[h!]
	\setlength{\tabcolsep}{15pt}
	\caption{Values of reflectance $R$, the incoherent sum of modal contributions $R_i$, the interference term $R_c$ in the cases of D1 and D2 in Fig. \ref{Fig4}(a), as well as their difference. } 
	\centering 
	\begin{tabular}{c   c   c   c} 
		\hline
		\hline
		\empty & $R$ & $R_i$ & $R_c$ \\ 
		\hline 
		D1 & 0.10 & 0.21 & -0.11  \\ 
		\hline
		D2 & 0.08 & 0.27 & -0.19 \\
		\hline
		D2$-$D1 & -0.02 & 0.06 & -0.08 \\ 
		\hline 
		\hline
	\end{tabular}
	\label{Stable3} 
\end{table}

\section{\label{sec:level7} Conclusion}

In summary, we present a numerical study on the minimum reflection channels in disordered waveguides and their modifications with the introduction of optical gain. 
A modal analysis reveals that the minimum reflection channel is formed by destructive interference of quasi-normal modes at the front surface of the random media. 
In most of the disordered configurations, the minimum reflection eigenvalue increases with gain, however, the minimum reflection channel can adjust its input wavefront to enhance the destructive interference of modes and slow down the growth of reflectance with gain. 
In some of the disordered waveguides, the differential amplification of the modes makes their destructive interference so effective that it dominates overs the amplitude growth of the modes, causing the reflectance to decrease with gain. 
Therefore, it is possible to further reduce light reflection from a strong scattering medium by adding optical gain to it. 
This counter-intuitive behavior illustrates the interplay between interference and amplification in a disordered system. \\

\section*{Acknowledgment}
We want to thank Arthur Goetschy, Li Ge, Alexander Cerjan, Allard P. Mosk and Willem Vos for helpful discussions. 
This work is funded by the US National Science Foundation under the Grant Nos. DMR-1205307 and ECCS-1068642.

\appendix*
\section{\label{sec:level1} Field decomposition by quasi-normal modes}
\begin{figure}[htbp!]
 	\centering
 	\includegraphics[width=\linewidth]{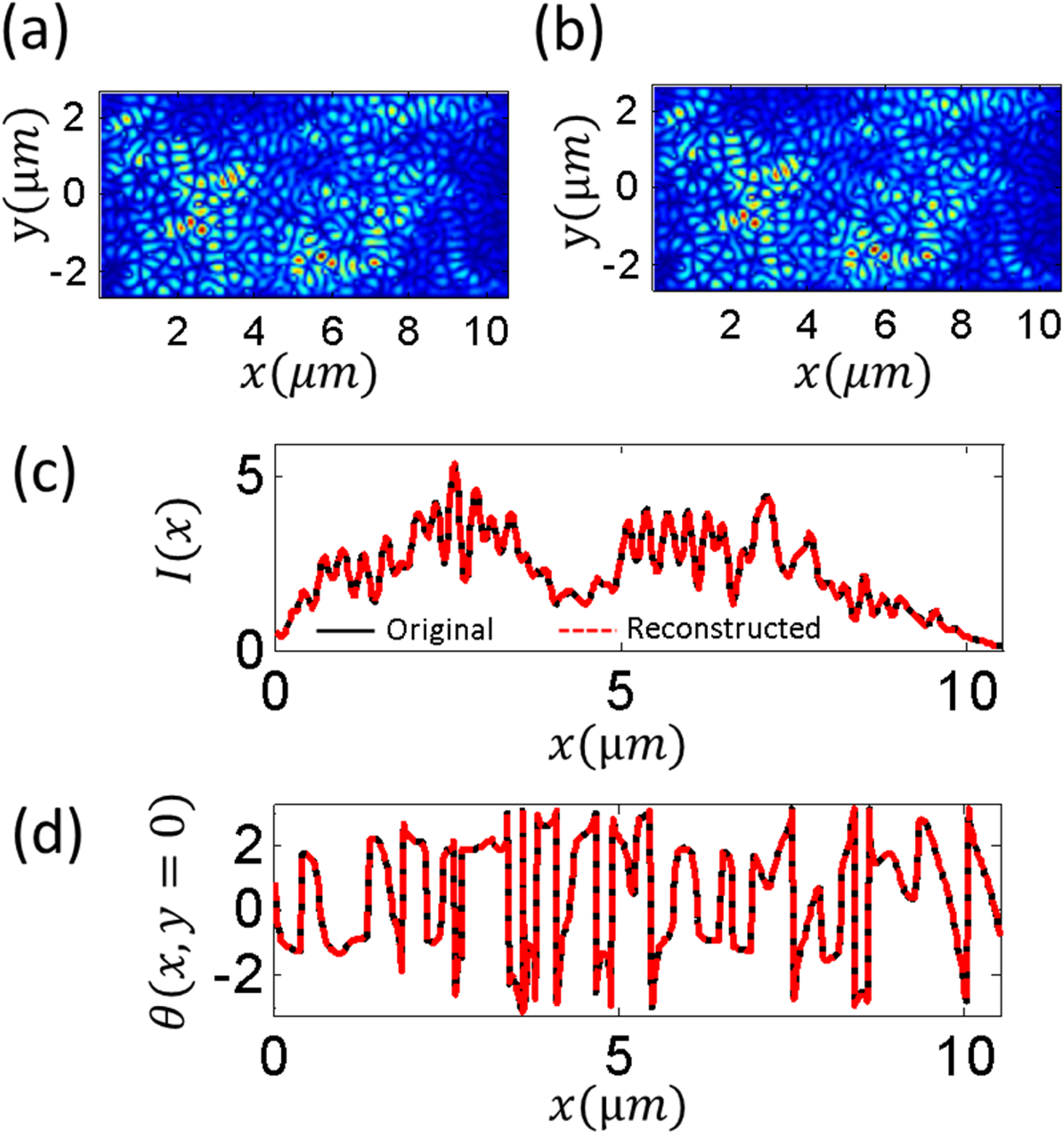}
 	\caption{(Color online) Field decomposition by quasi-normal modes. (a) The spatial distribution of electric field amplitude $|E_S(x,y)|$ inside the random waveguide with input wavefront corresponding to the minimum reflection channel without gain in Fig. \ref{Fig2}(a). (b) The reconstructed field distribution by modal decomposition $|E_C(x,y)|$. The cross-section integrated intensity $I(x) = \int |E(x,y)|^2 dy$ and phase of the electric field $E(x, y=0)$ between the original and reconstructed fields are shown in (c) and (d) respectively. The relative difference between the two field profiles is measured by $\int\int |E_C(x,y)-E_S(x,y)|^2dx\ dy/\int\int|E_S(x,y)|^2dx\ dy = 2.14\times 10^{-5}$. }
 	\label{Fig5}
 \end{figure}

To decompose the electric field distribution inside the random system by the quasi-normal modes, we use the finite-difference-frequency-domain (FDFD) method \cite{comsol_42a} to calculate the modes with incoming and outgoing wave boundary conditions. 
Figure \ref{Fig5}(a) shows the spatial distribution of the electric field amplitude $|E_S(x,y)|$ inside the disordered waveguide for the minimum reflection channel without gain in Fig. \ref{Fig2}(a). 
The electric field inside the random system is decomposed by the quasi-normal modes of the passive system. 
\begin{eqnarray}
E_C(x,y) = \sum_m a_m u_m(x,y) + \sum_m b_m v_m(x,y), \nonumber
\end{eqnarray}
where $u_m$ ($v_m$) is the $m$-th resonant mode with purely outgoing (incoming) wave boundary condition. 
The decomposition involves finding the coefficients $a_m$ and $b_m$ by fitting $E_C(x,y)$ to $E_S(x,y)$ using the nonlinear curve-fitting function \textit{lsqcurvefit} in MATLAB. 
After the decomposition, the reconstructed field profile in Fig. \ref{Fig5}(b) matches well the original one in Fig. \ref{Fig5}(a). 
The good agreement is further shown in the cross-section integrated intensity $I(x) = \int |E(x,y)|^2 dy$ and phase of the electric field $E(x, y=0)$ in Figs. \ref{Fig5}(c) and (d). 
The error of fitting, characterized by the relative difference between $E_S(x,y)$ and $E_C(x,y)$ is calculated by $\int\int |E_C(x,y)-E_S(x,y)|^2dx\ dy/\int\int|E_S(x,y)|^2dx\ dy = 2.14\times 10^{-5}$. 
 \pagebreak


\begin{thebibliography}{99}
\bibitem{dorokhov1}
O. N. Dorokhov, Solid State Commun. \textbf{44}, 915 (1982).
\bibitem{dorokhov2}
O. N. Dorokhov, Solid State Commun. \textbf{51}, 381 (1984).
\bibitem{Mello88}
P. A. Mello, P. Pereyra, and N. Kumar, Annals of Physics \textbf{181}, 290 (1988).
\bibitem{Nazarov94}
Y. V. Nazarov, Phys. Rev. Lett. \textbf{73}, 134 (1994).
\bibitem{vellekoop_OL}
I. M. Vellekoop and A. P. Mosk, Opt. Lett. \textbf{32}, 2309 (2007).
\bibitem{changhueiyang_NP}
Z. Yaqoob, D. Psaltis, M. S. Feld, and C. Yang, Nat. Photon. \textbf{2}, 110 (2008).
\bibitem{vellekoop_PRL}
I. M. Vellekoop and A. P. Mosk, Phys. Rev. Lett. \textbf{101}, 120601 (2008).
\bibitem{Pendry_view}
J. B. Pendry, Physics \textbf{1}, 20 (2008).
\bibitem{sebastien_PRL}
S. M. Popoff, G. Lerosey, R. Carminati, M. Fink, A. C. Boccara, and S. Gigan, Phys. Rev. Lett. \textbf{104}, 100601 (2010).
\bibitem{mosk_NP}
I. M. Vellekoop, A. Lagendijk, and A. P. Mosk, Nat. Photon. \textbf{4}, 320 (2010).
\bibitem{choi_PRB}
W. Choi, A. P. Mosk, Q. H. Park, and W. Choi, Phys. Rev. B \textbf{83}, 134207 (2011).
\bibitem{choi_OE}
W. Choi, Q. H. Park, and W. Choi, Opt. Express \textbf{20}, 20721 (2012).
\bibitem{shi_PRL12}
Z. Shi and A. Z. Genack, Phys. Rev. Lett. \textbf{108}, 043901 (2012).
\bibitem{choi_NP}
M. Kim, Y. Choi, C. Yoon, W. Choi, J. Kim, Q. Park, and W. Choi, Nat. Photon. \textbf{6}, 581 (2012).
\bibitem{Davy_OE13}
M. Davy, Z. Shi, J. Wang, and A. Z. Genack, Opt. Express \textbf{21}, 10367 (2013).
\bibitem{kim_OL13}
M. Kim, W. Choi, C. Yoon, G. H. Kim, and W. Choi , Opt. Lett. \textbf{38}, 2994 (2013).
\bibitem{sebastien13}
S. M. Popoff, A. Goetschy, S. F. Liew, A. D. Stone, and H. Cao, Phys. Rev. Lett. \textbf{112}, 133903 (2014).
\bibitem{liew_PRB14}
S. F. Liew, S. M. Popoff, A. P. Mosk, W. L. Vos, and H. Cao, Phys. Rev. B \textbf{89}, 224202 (2014).
\bibitem{Genack_NC15}
M. Davy, Z. Shi, J. Park, C. Tian, and A. Z. Genack, Nat. Commun. \textbf{6}, 6893 (2015).
\bibitem{sebastien_NC}
S. Popoff, G. Lerosey, M. Fink, A. C. Boccara, and S. Gigan, Nat. Commun. \textbf{1}, 81 (2010).
\bibitem{Lagendijk_PRL11}
J. Aulbach, B. Gjonaj, P. M. Johnson, A. P. Mosk, and A. Lagendijk, Phys. Rev. Lett. \textbf{106}, 103901 (2011).
\bibitem{Katz_NP}
O. Katz, E. Small, Y. Bromberg, and Y. Silberberg, Nat. Photon. \textbf{5}, 372 (2011).
\bibitem{Gigan_NP}
D. J. McCabe,	A. Tajalli,	D. R. Austin, P. Bondareff,	I. A. Walmsley,	S.n Gigan, and B\'{e}atrice Chatel, Nat. Commun. \textbf{2}, 447 (2011).
\bibitem{Mosk_NP12}
A. P. Mosk, A. Lagendijk, G. Lerosey, M. Fink, Nat. Photon. \textbf{6}, 283 (2012).
\bibitem{Park2012spectral}
 J. H. Park, C. H. Park, H. Yu, Y. H. Cho,, and Y. K. Park, Opt. Lett. \textbf{37}, 3261 (2012).
\bibitem{Small2012control}
E. Small, O. Katz, Y. Guan, and Y. Silberberg, Opt. Lett. \textbf{37}, 3429 (2012).
\bibitem{Piestun}
D. B. Conkey, A. M. Caravaca-Aguirre, and R. Piestun, Opt. Express \textbf{20}, 1733 (2012).
\bibitem{Mertz}
H. P. Paudel, C. Stockbridge, J. Mertz, and T. Bifano, Opt. Express \textbf{21}, 17299 (2013)
\bibitem{Yu_review}
H. Yu, J. Park, K. Lee, J. Yoon, K. Kim, S. Lee, and Y. K. Park, Curr. Appl. Phys. \textbf{15}, 632 (2015).
\bibitem{Kubby}
X. Tao, D. Bodington, M. Reinig, and J. Kubby, Opt. Express \textbf{23}, 14168 (2015).
\bibitem{Park_OptoGen}
J. Yoon, M. Lee, K. Lee, N. Kim, J. M. Kim, J. Park, C. Choi, W. D. Heo, Y. Park, arXiv:1502.04826 (2015). 
\bibitem{Park_OC15}
H. Yu, J.-H. Park, Y. K. Park,  Opt. Commun. \textbf{352}, 33 (2015). 
\bibitem{John_PRL84}
S. John, Phys. Rev. Lett. \textbf{53}, 2169 (1984).
\bibitem{Yurkevich_PRL94}
V. Freilikher, M. Pustilnik, and I. Yurkevich, Phys. Rev. Lett. \textbf{73}, 810 (1994).
\bibitem{zhang_PRB95}
Z.-Q. Zhang, Phys. Rev. B \textbf{52}, 7960 (1995).
\bibitem{beenakker_PhysA}
T. Sh. Misirpashaev, J. C. J. Paasschens, C. W. J. Beenakker, Physica A \textbf{236}, 189 (1997).
\bibitem{Heinrichs_PRB97}
J. Heinrichs, Phys. Rev. B \textbf{56}, 8674 (1997).
\bibitem{brouwer}
P. W. Brouwer, Phys. Rev. B \textbf{57}, 10526 (1998).
\bibitem{Datta_PRB99}
P. K. Datta, Phys. Rev. B \textbf{59}, 10980 (1999).
\bibitem{chabanov_Nat00}
A. A. Chabanov, M. Stoytchev, and A. Z. Genack, Nature \textbf{404}, 850 (2000).
\bibitem{Lisyansky_PRB01}
L. I. Deych, A. Yamilov, and A. A. Lisyansky, Phys. Rev. B \textbf{64}, 024201 (2001).
\bibitem{yamilov_OE13}
A. G. Yamilov and B. Payne, Opt. Express \textbf{21}, 11688 (2013).
\bibitem{Zhang_PRB13}
L.-Y. Zhao, C.-S. Tian, Z.-Q. Zhang, and X.-D. Zhang, Phys. Rev. B \textbf{88}, 155104 (2013).
\bibitem{liewOE15}
S. F. Liew and H. Cao, Opt. Express \textbf{23}, 11043 (2015).
\bibitem{Zyuzin_94}
A. Yu. Zyuzin, Europhys. Lett. \textbf{26}, 517 (1994).
\bibitem{Kumar_PRB94}
P. Pradhan and N. Kumar, Phys. Rev. B \textbf{50}, 9644 (1994).
\bibitem{Zyuzin_PRE95}
A. Yu. Zyuzin, Phys. Rev. E \textbf{51}, 5274 (1995).
\bibitem{Wiersma_PRL95}
D. S. Wiersma, M. P. van Albada, and A. Lagendijk, Phys. Rev. Lett. \textbf{75}, 1739 (1995).
\bibitem{beenakker_PRL96}
C. W. J. Beenakker, J. C. J. Paasschens, and P. W. Brouwer, Phys. Rev. Lett. \textbf{76}, 1368 (1996).
\bibitem{Burkov_JETP96}
A. A. Burkov and A. Yu. Zyuzin, JETP Lett. \textbf{63}, 878 (1996).
\bibitem{Burkov_PRB97}
A. A. Burkov and A. Yu. Zyuzin, Phys. Rev. B \textbf{55}, 5736 (1997).
\bibitem{Freilikher_PRB97}
V. Freilikher, M. Pustilnik, and I. Yurkevich, Phys. Rev. B \textbf{56}, 5974 (1997).
\bibitem{Jayannavar_PRB97}
S. K. Joshi and A. M. Jayannavar, Phys. Rev. B \textbf{56}, 12038 (1997).
\bibitem{soukoulis_PRB99}
X. Jiang and C. M. Soukoulis, Phys. Rev. B \textbf{59}, 6159 (1999).
\bibitem{Kumar_PRB00}
S. A. Ramakrishna and N. Kumar, Phys. Rev. B \textbf{61}, 3163 (2000). 
\bibitem{Kumar_PRB00v2}
S. A. Ramakrishna, E. Krishna Das, G. V. Vijayagovindan, and N. Kumar, Phys. Rev. B \textbf{62}, 256 (2000). 
\bibitem{Kumar_PhysE01}
N. Kumar, Physica E \textbf{9}, 356 (2001).
\bibitem{Yamilov_PRE04}
A. Yamilov and H. Cao, Phys. Rev. E \textbf{70}, 037603 (2004).
\bibitem{Yamilov_PRB05}
A. Yamilov, S. H. Chang, A. Burin, A. Taflove, and H. Cao, Phys. Rev. B \textbf{71}, 092201 (2005).
\bibitem{Yamilov_PRE06}
A. Yamilov and H. Cao, Phys. Rev. E \textbf{74}, 056609 (2006).
\bibitem{Yamilov_PRB10}
B. Payne,J. Andreasen, H. Cao, and A. Yamilov, Phys. Rev. B \textbf{82}, 104204 (2010).
\bibitem{CaoWRM}
H. Cao, Waves Random Media \textbf{13}, R1 (2003).
\bibitem{CaoJPA}
H. Cao, J. Phys. A: Math. Gen. \textbf{38} 10497 (2005).
\bibitem{wu_JOSAB07}
X. Wu, J. Andreasen, H. Cao, and A. Yamilov, J. Opt. Soc. Am. B \textbf{24}, A26 (2007).
\bibitem{comsol_42a}
http://www.comsol.com

%
\bibitem{genack_nature}
J. Wang and A. Z. Genack, Nature \textbf{471}, 345 (2011).
\bibitem{genack_JMP}
J. Wang, Z. Shi, M. Davy, and A. Z. Genack, Intern. J. Mod. Phys. \textbf{11}, 1 (2012). 
\bibitem{modalMakeup}
Z. Shi and A. Z. Genack, arXiv:1406.3673. 




\end{thebibliography}
\end{document}